\documentclass[12pt]{article}
\usepackage{graphicx}

\newcommand\pubnumber{}
\newcommand\pubdate{\today}

\def\torino{Department of Physics, Universit\`a di Torino\\
and\\
INFN, Sezione di Torino\\
via P.Giuria 1, I-10125 Torino, ITALY}
\def\support{\footnote{This work is supported by the research grant {\sl Theoretical Astroparticle Physics} number 2012CPPYP7 under the program PRIN 2012 funded by the Ministero dell'Istruzione, Universit\`a e della Ricerca (MIUR).}}

\def\Title#1{\begin{center} {\Large #1 } \end{center}}
\def\Author#1{\begin{center}{ \sc #1} \end{center}}
\def\Address#1{\begin{center}{ \it #1} \end{center}}

\newcommand\pubblock{\rightline{\begin{tabular}{l} \pubnumber\\
         \pubdate  \end{tabular}}}
\newenvironment{Abstract}{\begin{quotation}  }{\end{quotation}}
\newenvironment{Presented}{\begin{quotation} \begin{center} 
             PRESENTED AT\end{center}\bigskip 
      \begin{center}\begin{large}}{\end{large}\end{center} \end{quotation}}

\def\beq{\begin{equation}}
\def\eeq#1{\label{#1}\end{equation}}
\def\eeqn{\end{equation}}

\def\beqa{\begin{eqnarray}}
\def\eeqa#1{\label{#1}\end{eqnarray}}
\def\eeqan{\end{eqnarray}}

\let\bar=\overbar

\def\Dslash{\not{\hbox{\kern-4pt $D$}}}
\def\dslash{\not{\hbox{\kern-2pt $\del$}}}

\def\msb{{\bar{\ssstyle M \kern -1pt S}}}

\newcommand{\Neff}{{N}_{\mathrm{eff}}}
\newcommand{\mnu}{\Sigma\,m_\nu}
\newcommand{\e}[1]{\cdot 10^{#1}}
\newcommand{\mpcinv}{\, \mathrm{Mpc}^{-1}}
\newcommand{\pchip}{\texttt{PCHIP}}

\begin{document}
\begin{titlepage}
\pubblock

\vfill
\Title{Dark Radiation and Inflationary Freedom}
\vfill
\Author{Stefano Gariazzo\support}
\Address{\torino}
\vfill
\begin{Abstract}
We perform a cosmological analysis in which we allow
the primordial power spectrum of scalar perturbations to assume
a shape that is different with respect to the usual power-law,
arising from the simplest models of cosmological inflation.
We parametrize the primordial power spectrum with
a piecewise monotone cubic Hermite function and we use it
to investigate how the constraints on the various cosmological parameters change:
we find that the obtained limits are relaxed with respect to the power-law case,
if CMB polarization data are not included.
Moreover, the cosmological analyses provide us some indications
about the shape of the reconstructed primordial power spectrum, where
we notice possible features around $k\simeq0.002\mpcinv$ and $k\simeq0.0035\mpcinv$.
If confirmed in future analyses involving enhanced experimental data,
these features suggests that
the simplest cosmological inflation models may be incomplete.
\end{Abstract}
\vfill
\begin{Presented}
NuPhys2015, Prospects in Neutrino Physics

Barbican Centre, London, UK,  December 16--18, 2015
\end{Presented}
\vfill
\end{titlepage}

\def\thefootnote{\fnsymbol{footnote}}
\setcounter{footnote}{0}

\section{Introduction}

In the cosmological analyses,
one of the main assumptions about the early Universe
is the power-law form 
of the Primordial Power Spectrum (PPS),
that is predicted by the simplest models of inflation.
Deviations from the simplest inflationary models can in principle
lead to different shapes of the PPS with respect
to the power-law form.
Any cosmological analysis performed assuming a power-law PPS, in turn,
can give biased constraints on the cosmological parameters.

We study how the freedom in the PPS shape can affect
the limits on the cosmological parameters,
with particular interest on
the bounds on the presence in the early Universe of 
additional dark radiation.
Among the main candidates, the most studied in the literature
are axions
(not treated here, see Ref.~\cite{DiValentino:2016ikp})
or neutrinos.

\section{Method}
We base our analysis on a flat $\Lambda$CDM model, described by
the usual parameters:
the present-day physical CDM and baryon densities
$\Omega_{\rm cdm} h^2$
and $\Omega_{\rm b} h^2$, 
the angular sound horizon $\theta_{\rm s}$, 
the optical depth to reionization $\tau$.

We will study two properties of dark radiation,
considering the cases of three massive neutrinos or
of additional massless neutrinos.
To accommodate
the presence of dark radiation,
we extend then the $\Lambda$CDM model
varying 
the sum of the neutrino masses $\mnu$
or
the effective number of relativistic
degrees of freedom $\Neff$, respectively.

When we consider the standard \emph{power-law} (PL) PPS,
we describe it using the usual parameters $n_s$ and $\ln(10^{10}A_{s})$, 
respectively its spectral index and its amplitude.

Following \cite{Gariazzo:2014dla},
we parametrize the \emph{free} PPS of scalar perturbations
with a ``piecewise cubic Hermite interpolating polynomial''
or \pchip{}.
We use $N=12$ nodes to be interpolated with the \pchip{} function: 
ten equally spaced nodes in the range
$(k_2=0.001\mpcinv, k_{11}=0.35\mpcinv)$,
better constrained from the data, and
two nodes $k_1=5\e{-6}\mpcinv$ and $k_{12}=10\mpcinv$,
necessary to parametrize a non-constant behavior
at the outermost wavemodes.
The spectrum is described by 
$P_{s}(k)=P_0 \times \pchip(k, P_{s,j})$,
where
$P_0=2.2\e{-9}$ is an arbitrary normalization,
$P_{s,j}=P_s(k_j)/P_0$ and
$0.01\leq P_{s,j}\leq 10$.

For our analyses, we consider the following experimental data sets:
full CMB temperature plus polarization data at low multipoles only
(Planck~TT+lowP)
or
full temperature and polarization data
(Planck~TT,TE,EE+lowP) from the Planck 2015 release
\cite{Aghanim:2015xee},
and
the data on the matter power spectrum at different redshifts from
the WiggleZ Dark Energy Survey (MPkW) \cite{Parkinson:2012vd}.

\section{Results on $\Neff$ and $\mnu$}

\begin{figure}[t]
  \begin{center}
    \includegraphics[width=0.65\textwidth,page=1]{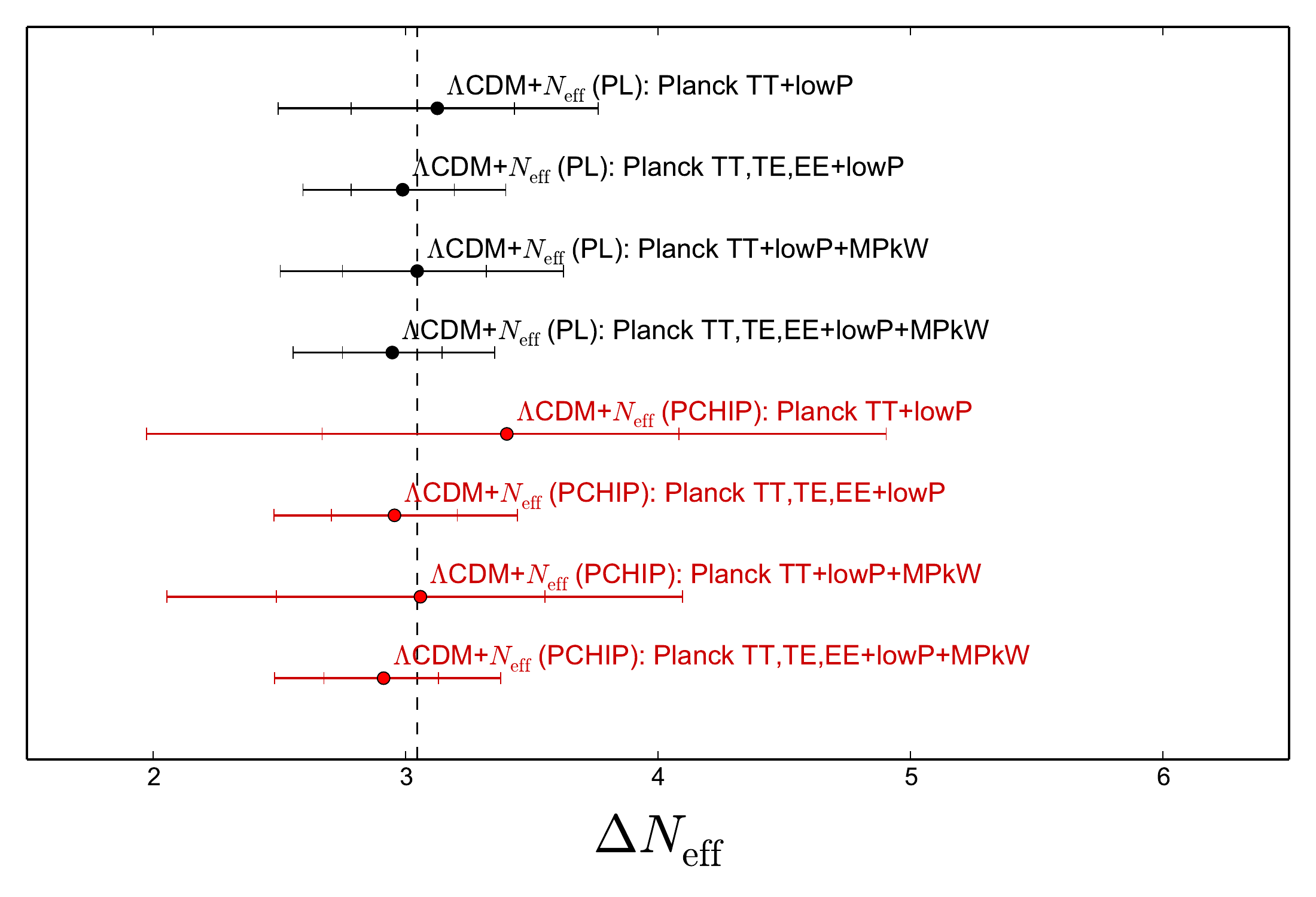}
    \caption{Limits on $\Neff$ at 1, 2 $\sigma$.
    Adapted from Ref.~\cite{DiValentino:2016ikp}.
    \label{fig:neff}}
  \end{center}
\end{figure}

The strongest degeneracies between the \pchip{} nodes and a cosmological
parameter appear for the effective number of relativistic species, $\Neff$.
A comparison of the $\Neff$ constraints
obtained using the power-law and the \pchip\ PPS
is reported in Fig.~\ref{fig:neff}.

The reason of the strong degeneracy is related to the effects
of varying $\Neff$ in cosmology.
When $\Neff$ is increased
with fixed matter-radiation equality
and matter-dark energy equality redshifts,
the Silk damping at small scales is enhanced
(see e.g.\ \cite{Gariazzo:2015rra}).
If one modifies the scalar PPS,
increasing it only at the scales interested by the 
enhanced Silk damping and not at the other scales,
the effects of a larger $\Neff$ can be canceled.
This is the reason for which the model with the free PPS,
considering Planck~TT+lowP data only,
allows high values of $\Neff$:
the PPS freedom allows to modify
the small scales without altering
the large scales, so that
a large $\Neff$ can be accommodated.
As expected,
the degeneracies between $\Neff$ and the nodes $P_{s,j}$
are stronger for the nodes at high wavemodes (small scales)
\cite{DiValentino:2016ikp}.

\begin{figure}[t]
  \begin{center}
  \includegraphics[width=0.65\textwidth,page=2]{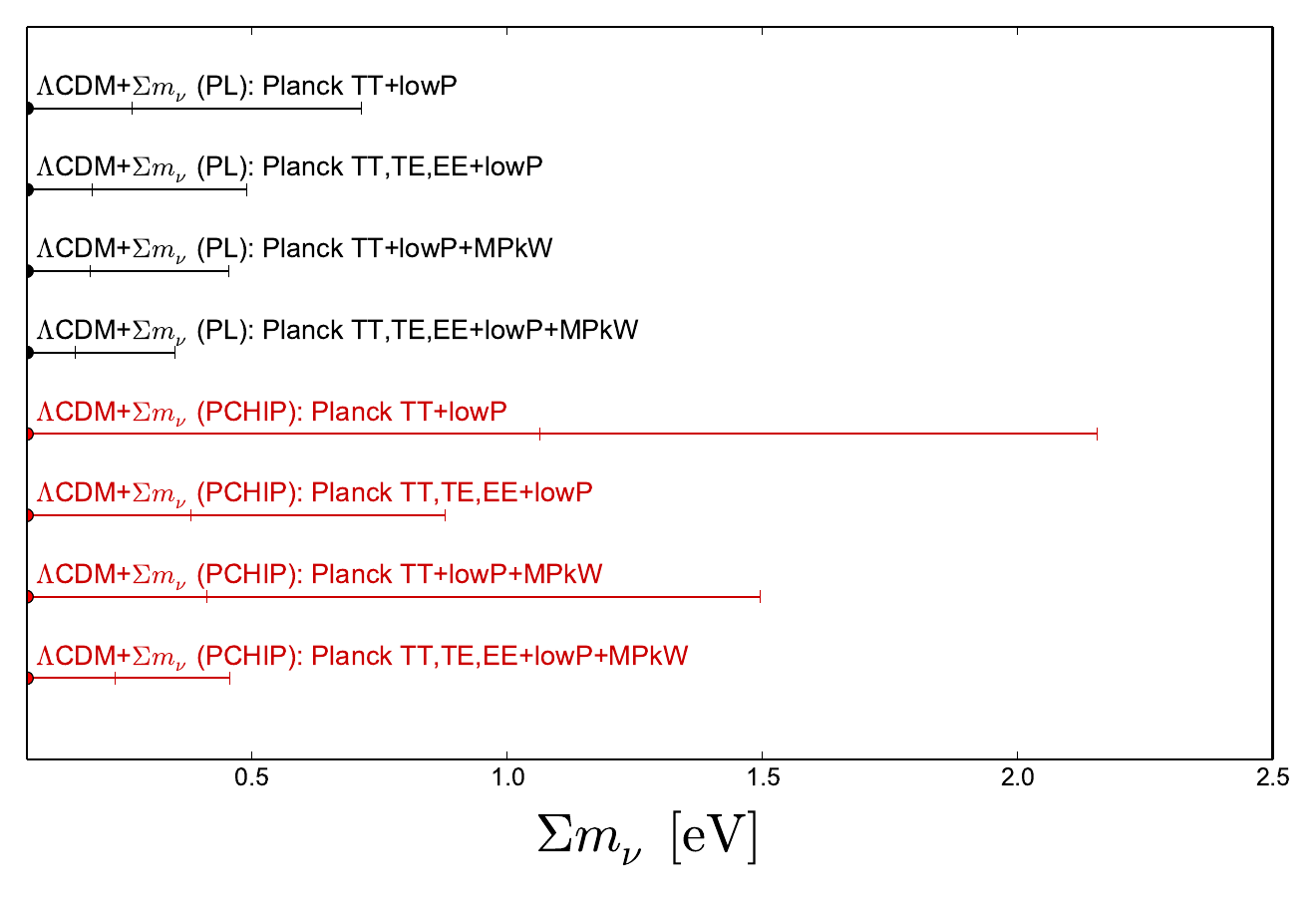}
    \caption{Limits on $\mnu$ at 1, 2 $\sigma$.
    Adapted from Ref.~\cite{DiValentino:2016ikp}.
    \label{fig:mnu}}
  \end{center}
\end{figure}

Cosmology can also constrain the absolute scale of neutrino masses.
The main effect of increasing the sum of the neutrino masses
is to obtain a change in the early and late ISW effects,
if the other cosmological parameters are changed
to fix the angular position of the CMB peaks.
Since the freedom in the PPS can compensate
the changes in the early and late ISW effects without altering
the other scales, the marginalized constraints on $\mnu$ are weakened
in the \pchip\ PPS case,
as it is possible to see in Fig.~\ref{fig:mnu}.

The inclusion of the CMB polarization at high multipoles
by Planck, however, prevents
the compensation between PPS parameters and neutrino properties, and
$\mnu$ and $\Neff$ are forced to be very close to the standard
values $\mnu\simeq0.06$~eV and $\Neff=3.046$, also with a free PPS.
This happens because the effect of a free PPS
on the TT, TE and EE spectra
is different from the effects
of increasing $\mnu$ or $\Neff$.

\section{PPS Results}
\begin{figure}[t]
  \begin{center}
    \includegraphics[width=0.65\textwidth,page=3]{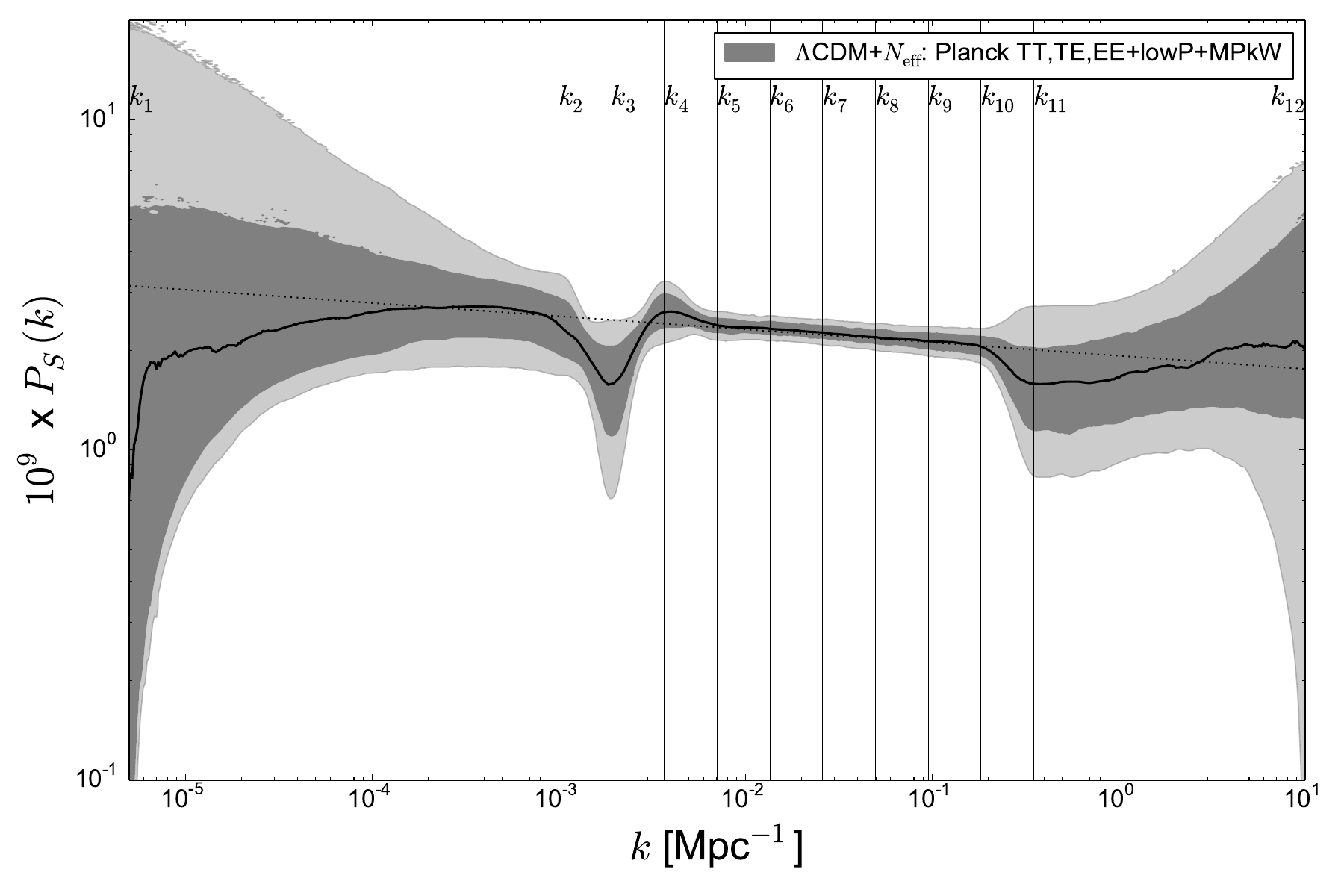}
  \end{center}
\caption{1, 2 $\sigma$ bounds on the constrained PPS
compared with the best-fit PL PPS.
From Ref.~\cite{DiValentino:2016ikp}.
\label{fig:pps}}
\end{figure}

An helpful way to visualize how the free PPS
is constrained by data in our model 
is to plot the marginalized constraints on the PPS shape,
as in Fig.~\ref{fig:pps}.
These are obtained marginalizing over the obtained PPS for each different
value of $k$ independently.

The reconstructed PPS has several interesting properties.
First of all, the least constrained nodes are in $k=5\e{-6}\mpcinv$
and $k=10\mpcinv$, as expected due to the absence of data at these wavemodes.
The nodes from $k\simeq0.007\mpcinv$ to $k\simeq0.2\mpcinv$, instead,
are the best constrained.
In this region the \pchip~PPS is in perfect agreement with the power-law PPS.
Two features appear at small wavemodes:
there are a small dip ($\simeq2\sigma$)
at $k\simeq0.002\mpcinv$, corresponding to 
the dip at $\ell\simeq25$ in the CMB temperature spectrum,
and a small bump ($\simeq1\sigma$) at $k\simeq0.0035\mpcinv$,
corresponding to the small bump at $\ell\simeq40$ in the CMB spectrum.

\section{Conclusions}
We noticed the presence of possible degeneracies
between the PPS of scalar perturbations and the parameters
that describe the dark radiation properties.
As a consequence, the limits 
on the effective number of relativistic particles $\Neff$
and on the total neutrino mass $\mnu$
from the CMB temperature data are relaxed.
The CMB polarization, however, helps
to break the degeneracies between the dark radiation parameters
and those related to the PPS of scalar perturbations.

The analyses show that possible features in the PPS exist,
in particular at large scales.
These may suggest that some new mechanism for inflation must be present,
in order to introduce a scale dependency
in the initial power spectrum of scalar perturbations,
as it is observed in the CMB spectrum.


\begin{thebibliography}{99}

\bibitem{DiValentino:2016ikp}
  E.~Di Valentino, S.~Gariazzo, M.~Gerbino, E.~Giusarma and O.~Mena,
  arXiv:1601.07557 [astro-ph.CO].
  
\bibitem{Gariazzo:2014dla}
  S.~Gariazzo, C.~Giunti and M.~Laveder,
  JCAP {\bf 1504} (2015) 04,  023
  doi:10.1088/1475-7516/2015/04/023
  [arXiv:1412.7405 [astro-ph.CO]].

\bibitem{Aghanim:2015xee}
  N.~Aghanim {\it et al.} [Planck Collaboration],
  [arXiv:1507.02704 [astro-ph.CO]].

\bibitem{Parkinson:2012vd}
  D.~Parkinson {\it et al.},
  Phys.\ Rev.\ D {\bf 86} (2012) 103518
  doi:10.1103/PhysRevD.86.103518
  [arXiv:1210.2130 [astro-ph.CO]].
  
\bibitem{Gariazzo:2015rra}
  S.~Gariazzo, C.~Giunti, M.~Laveder, Y.~F.~Li and E.~M.~Zavanin,
  J.\ Phys.\ G {\bf 43} (2016) 033001
  doi:10.1088/0954-3899/43/3/033001
  [arXiv:1507.08204 [hep-ph]].
 
\end{thebibliography}
\end{document}